\documentstyle[prd,aps]{revtex}

\newcommand{\bea}{\begin{eqnarray}}
\newcommand{\eea}{\end{eqnarray}}
 
\begin{document}

\draft

\title{Relativistic Hydrodynamic Cosmological Perturbations}
\author{Jai-chan Hwang${}^{(a,c)}$ and Hyerim Noh${}^{(b,c)}$}
\address{${}^{(a)}$ Department of Astronomy and Atmospheric Sciences,
                    Kyungpook National University, Taegu, Korea \\
         ${}^{(b)}$ Korea Astronomy Observatory,
                    San 36-1, Whaam-dong, Yusung-gu, Daejon, Korea \\
         ${}^{(c)}$ Max-Planck-Institut f\"ur Astrophysik,
                    Karl-Schwarzschild-Str. 1,
                    85740 Garching bei M\"unchen, Germany}
\date{\today}
\maketitle

\begin{abstract}

Relativistic cosmological perturbation analyses can be made
based on several different fundamental gauge conditions.
In the pressureless limit the variables in certain gauge conditions
show the correct Newtonian behaviors.
Considering the general curvature ($K$) and the cosmological constant 
($\Lambda$) in the background medium,
the perturbed density in the comoving gauge, and the perturbed velocity 
and the perturbed potential in the zero-shear gauge show the same behavior 
as the Newtonian ones {\it in general scales}.
In the first part, we elaborate these Newtonian correspondences.
In the second part, using the identified gauge-invariant variables with
correct Newtonian correspondences, we present the relativistic results with 
general pressures in the background and perturbation. 
We present the general {\it super-sound-horizon scale solutions} 
of the above mentioned variables valid for general $K$, 
$\Lambda$, and generally evolving equation of state.
We show that, for vanishing $K$, the super-sound-horizon scale evolution is 
characterised by {\it a conserved variable} which is the perturbed  
three-space curvature in the comoving gauge.
We also present equations for the multi-component hydrodynamic situation 
and for the rotation and gravitational wave.


\end{abstract}

\section{Introduction} 
                                    \label{sec:Introduction}

The analysis of gravitational instability in the expanding universe
model was first presented by Lifshitz in 1946 in a general relativistic 
context \cite{Lifshitz}.
Historically, the much simpler, and in hindsight, more intuitive 
Newtonian study followed later \cite{Bonnor}.
The pioneering study by Lifshitz is based on a gauge choice which is
commonly called the synchronous gauge. 
As the later studies have shown, the synchronous gauge is only 
one way of fixing the gauge freedom out of several available
fundamental gauge conditions
\cite{Harrison,Nariai,Bardeen-1980,KS,Reviews,PRW}.
As will be summarized in the following, out of the several gauge conditions
only the synchronous gauge fails to fix the gauge mode completely and 
thus often needs more involved algebra.
As long as one is careful of the algebra this gauge choice does 
not cause any kind of intrinsic problem; there exist, however, some persisting
algebraic errors in the literature, see Sec. \ref{sec:Discussion}.
The gauge condition which turns out to be especially suitable for
handling the perturbed density is the comoving gauge used first by
Nariai in 1969 \cite{Nariai}.
Since the comoving gauge condition completely fixes the gauge transformation
property, the variables in this gauge can be equivalently considered as
gauge-invariant ones.
As mentioned, there exist several such fundamental gauge conditions
each of which completely fixes the gauge transformation properties.
One of such gauge conditions suitable for handling the gravitational
potential and the velocity perturbations is the zero-shear gauge
used first by Harrison in 1967 \cite{Harrison}.
The variables in such gauge conditions are equivalently gauge-invariant.
Using the gauge freedom as an advantage for handling the problem
was emphasized by Bardeen in 1988 \cite{Bardeen-1988}.
In order to use the gauge choice as an advantage a {\it gauge ready method}
was proposed in \cite{PRW} which will be adopted in the 
following, see Sec. \ref{sec:Relativistic}.

The variables which characterize the self gravitating Newtonian fluid flow are
the density, the velocity and the gravitational potential 
(the pressure is specified by an equation of state),
whereas, the relativistic flow may be characterized by various components
of the metric (and consequent curvatures) and the energy-momentum tensor.
Since the relativistic gravity theory is a constrained system we have the 
freedom of imposing certain conditions on the metric or the energy-momentum 
tensor as coordinate conditions.  
In the perturbation analyses the freedom arises because we need to introduce
a fictitious background system in order to describe the physical perturbed
system.
The correspondence of a given spacetime point between the perturbed spacetime
and the ficticious background one could be made with certain degrees 
of freedom.
This freedom can be fixed by suitable conditions (the gauge conditions)
based on the spacetime coordinate transformation.
Studies in the literature show that a certain variable in a certain gauge
condition correctly reproduces the corresponding Newtonian behavior.
Although the perturbed density in the comoving gauge shows the Newtonian
behavior, the perturbed potential and the perturbed velocity in the
same gauge do not behave like the Newtonian ones; for example, in the comoving
gauge the perturbed velocity vanishes by the coordinate (gauge) condition.
It is known that both the perturbed potential and the perturbed velocity
in the zero-shear gauge correctly behave like the corresponding
Newtonian variables \cite{Bardeen-1980}.

In the first part (Sec. \ref{sec:Six-gauges}) we will elaborate establishing 
such correspondences between the relativistic and Newtonian perturbed 
variables.
Our previous work on this subject is presented in \cite{MDE,Hyun};
compared with \cite{MDE} in this work we will explicitly compare 
the relativistic equations for the perturbed density, potential and velocity 
variables in several available gauge conditions with the
ones in the Newtonian system.
We will include both the background spatial curvature ($K$) 
and the cosmological constant ($\Lambda$).
In the second part (Sec. \ref{sec:Hydrodynamics}), using the variables 
with correct Newtonian correspondences,
we will extend our relativistic results to the situations
with general pressures in the background and perturbations.
We will present the relativistic equations satisfied by the gauge-invariant
variables and will derive the {\it general solutions} valid in the 
super-sound-horizon scale (i.e., larger than Jeans scale)
considering both $K$ and $\Lambda$, and the generally evolving
ideal fluid equation of state $p = p(\mu)$.

Section \ref{sec:review} is a review of our previous work
displaying the basic equations in both Newtonian and relativistic contexts
and summarizing our strategy for handling the equations. 
In Sec. \ref{sec:Six-gauges} we consider a pressureless limit of the
relativistic equations.
We derive the equations for the perturbed density, the perturbed potential 
and the perturbed velocity in several different fundamental gauge conditions.
By comparing these relativistic equations in several gauges with
the Newtonian ones we {\it identify} the gauge conditions which reproduce 
the correct Newtonian behavior for certain variables in general scales.
In Sec. \ref{sec:Hydrodynamics} we present the fully relativistic equations 
for the identified gauge-invariant variables with correct Newtonian limits, 
now, considering the general pressures in the background and perturbations.
We derive the general large-scale solutions valid for general $K$ and $\Lambda$
in an ideal fluid medium, which are thus valid for general equation of 
state of the 
form $p = p(\mu)$, but with negligible entropic and anisotropic pressures.
The solutions are valid in the super-sound-horizon scale, thus are valid 
effectively in all scales in the matter dominated era where the
sound-horizon (Jeans scale) is negligible.
We discuss several quantities which are conserved in the large-scale
under the changing background equation of state.
These are the perturbed three-space curvature in several gauge conditions,
and in particular, we find the three-space curvature perturbation
in the comoving gauge shows a distinguished behavior:
for $K = 0$ (but for general $\Lambda$), it is conserved in a
super-sound-horizon scale independently of the changing equation of state.
{}For completeness we also summarize the case with multiple fluid
components in Sec. \ref{sec:Multi}, and cases of the rotation and
the gravitational wave in Sec. \ref{sec:rot-GW}.
Sec. \ref{sec:Hydrodynamics} is the highlight of the present work.
Sec. \ref{sec:Discussion} is a discussion.
We set $c \equiv 1$.

\section{Basic Equations and Strategy} 
                                    \label{sec:review}

\subsection{Newtonian Cosmological Perturbations} 
                                    \label{sec:Newtonian}

Since the Newtonian perturbation analysis in the expanding medium is well known
we begin by summarizing the basic equations \cite{MDE,Weinberg,Peebles-1980}.
The background evolution is governed by
\bea
   & & H^2 = {8 \pi G \over 3} \varrho - {K \over a^2} + {\Lambda \over 3}, 
       \quad \varrho \propto a^{-3},
   \label{Newt-BG}
\eea
where we allowed the general curvature (total energy) 
and the cosmological constant; $a(t)$ is a cosmic scale factor,
$H (t) \equiv \dot a / a$, and $\varrho (t)$ is the mass density. 
In Newtonian theory the cosmological constant can be introduced in the
Poisson equation {\it by hand} as 
$\nabla^2 \Phi = 4 \pi G \varrho - \Lambda$.
Perturbed parts of the mass conservation, the momentum conservation
and the Poisson's equations are [see Eqs. (43,46) in \cite{MDE}]: 
\bea
   & & \delta \dot \varrho + 3 H \delta \varrho = - {k \over a} \varrho
       \delta v, \quad
       \delta \dot v + H \delta v = {k \over a} \delta \Phi, \quad
       - {k^2 \over a^2} \delta \Phi = 4 \pi G \delta \varrho,
   \label{Newt-eqs} 
\eea
where $\delta \varrho ({\bf k}, t)$, $\delta v ({\bf k}, t)$ and
$\delta \Phi ({\bf k},t)$ are the Fourier modes of the perturbed mass density, 
velocity, and gravitational potential, respectively 
[see Eqs. (44,45) in \cite{MDE}];
$k$ is a comoving wave number with $\nabla^2 \rightarrow - k^2/a^2$.
[For linear perturbations the same forms of equations are valid in the 
configuration and the Fourier spaces.
Thus, without causing any confusion, we often ignore distinguishing the
Fourier space from the configuration space by an additional subindex.]
Equation (\ref{Newt-eqs}) can be arranged into the closed form equations
for $\delta$ ($\equiv \delta \varrho / \varrho$), $\delta v$ and
$\delta \Phi$ as:
\bea
   & & \ddot \delta + 2 H \dot \delta - 4 \pi G \varrho \delta 
       = {1 \over a^2 H} \left[ a^2 H^2 \left( {\delta \over H} \right)^\cdot
       \right]^\cdot = 0,
   \label{Newt-delta-eq} \\
   & & \delta \ddot v + 3 H \delta \dot v + \left( \dot H + 2 H^2 
       - 4 \pi G \varrho \right) \delta v = 0,
   \label{Newt-v-eq} \\
   & & \delta \ddot \Phi + 4 H \delta \dot \Phi
       + \left( \dot H + 3 H^2 - 4 \pi G \varrho \right) \delta \Phi 
       = {1 \over a^3 H} \left[ a^2 H^2 
       \left( {a \over H} \delta \Phi \right)^\cdot \right]^\cdot = 0.
   \label{Newt-Phi-eq} 
\eea
We note that Eqs. (\ref{Newt-eqs}-\ref{Newt-Phi-eq}) are
valid for general $K$ and $\Lambda$.
Although redundant, we presented these equations for later comparison
with the relativistic results.
The general solutions for $\delta$, $\delta v$ and $\delta \Phi$
immediately follow as (see also Table 1 in \cite{MDE}):
\bea
   & & \delta ({\bf k},t) = k^2 \left[ 
       H C ({\bf k}) \int^t_0 {dt \over \dot a^2} 
       + {H \over 4 \pi G \varrho a^3} d ({\bf k}) \right], 
   \label{delta-N} \\
   & & \delta v ({\bf k}, t) = - \left[ {k \over aH} C ({\bf k})
       \left( 1 + a^2 H \dot H \int^t_0 {dt \over \dot a^2} \right)
       + {k \dot H \over 4 \pi G \varrho a^2} d ({\bf k}) \right],
   \label{v-N} \\
   & & \delta \Phi ({\bf k}, t) = - \left[ 4 \pi G \varrho a^2 H C ({\bf k}) 
       \int^t_0 {dt \over \dot a^2} + {H \over a} d ({\bf k}) \right].
   \label{Phi-N}
\eea
The coefficients $C ({\bf k})$ and $d ({\bf k})$ are two integration constants, 
and indicate the relatively growing and decaying solutions, respectively;
the coefficients are matched in accordance with the solutions with a general
pressure in Eqs. (\ref{varphi_chi-sol}-\ref{v_chi-sol}) using 
Eq. (\ref{Correspondences-2})
(different dummy variables for $t$ in the integrands are assumed).
In the literature we can find various analytic form expressions of the 
above solutions in diverse situations, 
see \cite{Lifshitz,Harrison,Nariai,Weinberg,Peebles-1980,MDE-sols}.

\subsection{Relativistic Cosmological Perturbations} 
                                    \label{sec:Relativistic}

In the relativistic theory the fundamental variables are the metric and
the energy-momentum tensor.
As a way of summarizing our notation we present our convention of
the metric and the energy-momentum tensor.
As the metric we consider a spatially homogeneous and isotropic spacetime
with the most general perturbation
\bea
   d s^2 = - a^2 \left( 1 + 2 \alpha \right) d \eta^2
       - a^2 \left( \beta_{,\alpha} +B_\alpha \right) d \eta d x^\alpha
       + a^2 \left[ g_{\alpha\beta}^{(3)} \left( 1 + 2 \varphi \right)
       + 2 \gamma_{|\alpha\beta} + 2 C_{(\alpha|\beta)} + 2 C_{\alpha\beta}
       \right] d x^\alpha d x^\beta,
   \label{metric} 
\eea
where $0 = \eta$ with $dt \equiv a d \eta$,
Indices $\alpha$, $\beta$ $\dots$ are spatial ones and
$g^{(3)}_{\alpha\beta}$ is the three-space metric of the homogeneous and
isotropic space.
$\alpha ({\bf x}, t)$, $\beta ({\bf x}, t)$, $\gamma ({\bf x}, t)$,
and $\varphi ({\bf x}, t)$ indicate the scalar-type structure with
four degrees of freedom.
The transverse $B_\alpha ({\bf x}, t)$ and $C_\alpha ({\bf x}, t)$
indicate the vector-type structure with four degrees of freedom.
The transverse-tracefree $C_{\alpha\beta} ({\bf x}, t)$
indicates the tensor-type structure with two degrees of freedom.
The energy-momentum tensor is decomposed as:
\bea
   & & T^0_0 \equiv - \bar \mu - \delta \mu, \quad
       T^0_\alpha \equiv - {1 \over k} 
       \left( \bar \mu + \bar p \right) v_{,\alpha}
       + \delta T^{(v)0}_{\;\;\;\;\;\alpha},
   \nonumber \\
   & & T^\alpha_\beta
       \equiv \left( \bar p + \delta p \right) \delta^\alpha_\beta 
       + {1 \over a^2} \left( \nabla^\alpha \nabla_\beta
       - {1 \over 3} \Delta \delta^\alpha_\beta \right) \sigma
       + \delta T^{(v)\alpha}_{\;\;\;\;\;\beta}
       + \delta T^{(t)\alpha}_{\;\;\;\;\;\beta},
   \label{Tab}
\eea
where $\mu ({\bf k}, t) \equiv \bar \mu (t) + \delta \mu ({\bf k}, t)$ and 
$p \equiv \bar p + \delta p$ are the energy density and the pressure, 
respectively; an overbar indicates a background order quantity and will 
be ignored unless necessary.
$v$ is a frame-independent velocity related perturbed order variable
and $\sigma$ is an anisotropic pressure.
Situation with multiple fluids will be considered in Sec. \ref{sec:Multi}.
$\delta T^{(v)0}_{\;\;\;\;\;\alpha}$, $\delta T^{(v)\alpha}_{\;\;\;\;\;\beta}$,
and $\delta T^{(t)\alpha}_{\;\;\;\;\;\beta}$ are vector and tensor-type
perturbed order energy-momentum tensor.
All the spatial tensor indices and the vertical bar are based on 
$g^{(3)}_{\alpha\beta}$ as the metric.
The three types of structures are related to the density condensation,
the rotation, and the gravitational wave, respectively.
Due to the high symmetry in the background the three types of perturbations
decouple from each other to the linear order.
Since these three types of structures evolve independently, we can handle
them separately.
Evolutions of the rotation and the gravitational wave will be discussed
in Sec. \ref{sec:rot-GW}.

{}For the background equations we have [Eq. (6) in \cite{Ideal}]:
\bea
   H^2 = {8 \pi G \over 3} \mu - {K \over a^2} + {\Lambda \over 3}, \quad
       \dot H = - 4 \pi G ( \mu + p ) + {K \over a^2}, \quad
       \dot \mu = - 3 H \left( \mu + p \right). 
   \label{BG-eqs} 
\eea
These equations follow from $G^0_0$ and $G^\alpha_\alpha - 3 G^0_0$
components of Einstein equations and $T^b_{0;b} = 0$, respectively;
the third equation follows from the first two equations.
{}For $p = 0$ and replacing $\mu$ with $\varrho$ Eq. (\ref{BG-eqs}) 
reduces to Eq. (\ref{Newt-BG}).
The relativistic cosmological perturbation equations without fixing
the temporal gauge condition, thus in a gauge ready form,
were derived in Eq. (22-28) of \cite{PRW}
[see also Eqs. (8-14) in \cite{Ideal}; we use 
$\delta \equiv \delta \mu /\mu$ and $v \equiv - (k/a) \Psi/(\mu + p)$]:
\bea
   & & \kappa \equiv 3 \left( - \dot \varphi + H \alpha \right)
       + {k^2 \over a^2} \chi,
   \label{eq1} \\
   & & - {k^2 - 3K \over a^2} \varphi + H \kappa = - 4 \pi G \mu \delta,
   \label{eq2} \\
   & & \kappa - {k^2 - 3K \over a^2} \chi = 12 \pi G \left( \mu + p \right)
       {a \over k} v,
   \label{eq3} \\
   & & \dot \chi + H \chi - \alpha - \varphi = 8 \pi G \sigma,
   \label{eq4} \\
   & & \dot \kappa + 2 H \kappa = \left( {k^2 \over a^2} - 3 \dot H \right) 
       \alpha + 4 \pi G \left( 1 + 3 c_s^2 \right) \mu \delta 
       + 12 \pi G e,
   \label{eq5} \\
   & & \dot \delta + 3 H \left( c_s^2 - {\rm w} \right) \delta 
       + 3 H { e \over \mu } = \left( 1 + {\rm w} \right)
       \left( \kappa - 3 H \alpha - {k \over a} v \right),
   \label{eq6} \\
   & & \dot v + \left( 1 - 3 c_s^2 \right) H v = {k \over a} \alpha
       + {k \over a \left( 1 + {\rm w} \right) } \left( c_s^2 \delta
       + {e \over \mu} - {2\over 3} {k^2 - 3K \over a^2} {\sigma \over \mu}
       \right),
   \label{eq7} 
\eea
where we introduced a spatially gauge-invariant combination
$\chi \equiv a ( \beta + a \dot \gamma)$.
According to their origins, Eqs. (\ref{eq1}-\ref{eq7}) can be called: 
the definition of $\kappa$, 
ADM energy constraint ($G^0_0$ component of Einstein equations), 
momentum constraint ($G^0_\alpha$ component), 
ADM propagation 
($G^\alpha_\beta - {1 \over 3} \delta^\alpha_\beta G^\gamma_\gamma$ component),
Raychaudhuri ($G^\alpha_\alpha - G^0_0$ component), 
energy conservation ($T^b_{0;b} = 0$),
and momentum conservation ($T^b_{\alpha;b} = 0$) equations, respectively.
The perturbed metric variables $\varphi ({\bf k}, t)$, $\kappa ({\bf k},t)$,
$\chi ({\bf k}, t)$ and $\alpha ({\bf k}, t)$ are the perturbed part of
the three space curvature, expansion, shear, and the lapse function, 
respectively; meanings for $\varphi$, $\kappa$, and $\chi$ are based on
the normal frame vector field.
The isotropic pressure is decomposed as
\bea
   \delta p ({\bf k}, t) \equiv c_s^2 (t) \delta \mu ({\bf k}, t) 
       + e ({\bf k}, t), \quad
       c_s^2 \equiv {\dot p \over \dot \mu}, \quad
       {\rm w} (t) \equiv {p \over \mu}.
   \label{p-decomposition}
\eea

Under the gauge transformation $\tilde x^a = x^a + \xi^a$, the variables
transform as (see Sec. 2.2 in \cite{PRW}):
\bea
   & & \tilde \alpha = \alpha - \dot \xi^t, \quad
       \tilde \varphi = \varphi - H \xi^t, \quad
       \tilde \chi = \chi - \xi^t, \quad
       \tilde \kappa = \kappa 
       + \left( 3 \dot H - {k^2 \over a^2} \right) \xi^t,
   \nonumber \\
   & & \tilde v = v - {k \over a} \xi^t, \quad
       \tilde \delta = \delta + 3 \left( 1 + {\rm w} \right) H \xi^t,
   \label{GT}
\eea
where $\xi^t \equiv e^s \xi^\eta$, and $e$ and $\sigma$ are 
gauge-invariant.
[Due to the spatial homogeneity of the background, the effect 
 from the spatial gauge transformation has been trivially handled; 
 $\chi$ and $v$ are the spatially gauge-invariant combinations of 
 the variables, and the other metric and fluid variables 
 are naturally spatially gauge-invariant \cite{Bardeen-1988}.]
Thus, the perturbed variables in Eqs. (\ref{eq1}-\ref{eq7})
are {\it designed} so that any one of the following conditions
can be used to fix the freedom based on the temporal gauge transformation:
$\alpha \equiv 0$ (the synchronous gauge), 
$\varphi \equiv 0$ (the uniform-curvature gauge),
$\chi \equiv 0$ (the zero-shear gauge), 
$\kappa \equiv 0$ (the uniform-expansion gauge), 
$v/k \equiv 0$ (the comoving gauge), 
and $\delta \equiv 0$ (the uniform-density gauge).
These gauge conditions contain most of the gauge conditions used in the
literature concerning the cosmological hydrodynamic perturbation in a single
component situation; for the multiple component case see Eq. (\ref{Gauges-2}).
Of course, we can make an infinite number of different linear combinations 
of these gauge conditions each of which is also suitable for the temporal 
gauge fixing condition.
Our reason for choosing these conditions as fundamental is partly based 
on conventional use, but apparently, as the names of the conditions indicate,
the gauge conditions also have reasonable meanings based on the metric and 
the energy-momentum tensors.

Each one of these gauge conditions, except for the synchronous gauge, 
fixes the temporal gauge transformation property completely.
Thus, each variable in these five gauge conditions 
is equivalent to a corresponding gauge-invariant combination
(of the variable concerned and the variable used in the gauge condition).
We proposed a convenient way of writing the gauge-invariant variables 
\cite{PRW}.
{}For example, we let
\bea
   & & \delta_v \equiv \delta + 3 (1 + {\rm w}) {aH \over k} v
       \equiv 3 (1 + {\rm w}) {a H \over k} v_\delta, \quad
       \varphi_\chi \equiv \varphi - H \chi \equiv - H \chi_\varphi, \quad
       v_\chi \equiv v - {k \over a} \chi \equiv - {k \over a} \chi_v.
   \label{GI-combinations}
\eea
The variables $\delta_v$, $v_\chi$ and $\varphi_\chi$ are gauge-invariant 
combinations; $\delta_v$ becomes $\delta$ in the comoving gauge which takes
$v/k=0$ as the gauge condition, etc.
In this manner we can systematically construct the corresponding 
gauge-invariant combination for any variable based on a gauge condition which 
fixes the temporal gauge transformation property completely.
One variable evaluated in different gauges can be considered as different 
variables, and they show different behaviors in general.
In this sense, except for the synchronous gauge, the variables in the 
rest of five gauges can be considered as the gauge-invariant variables.
[The variables with a subindex $\alpha$ are not gauge-invariant, 
because those are equivalent to variables in the synchronous gauge.]
Although $\delta_v$ is a gauge-invariant variable which becomes
$\delta$ in the comoving gauge, $\delta_v$ itself 
can be evaluated in any gauge with the same value.

Equations (\ref{eq1}-\ref{eq7}) are the basic set of gauge-ready form
equations for hydrodynamic perturbations proposed in \cite{PRW}: 
``The moral is that one should work in the gauge that is mathematically
most convenient for the problem at hand'', \cite{Bardeen-1988}.
Accordingly, in order to use the available gauge conditions as the
advantage, Eqs. (\ref{eq1}-\ref{eq7}) are designed 
for easy implementation of the fundamental gauge conditions.
Using this {\it gauge-ready formulation}, complete sets of solutions for the 
six different gauge conditions are presented in an ideal fluid case 
\cite{Ideal}, and in a pressureless medium \cite{MDE}.
Equations (\ref{BG-eqs}-\ref{p-decomposition}) include general pressures which 
may account for the nonequilibrium or dissipative effects in hydrodynamic 
flows in the cosmological context with general $K$ and $\Lambda$.
The equations are expressed in general forms so that the fluid quantities 
can represent successfully the effects of the scalar field and 
classes of generalized gravity theories \cite{PRW}.
The applications of the gauge-ready formalism to the minimally coupled scalar
field and to the generalized gravity theories have been made in \cite{MSF-GGT}.

\section{Newtonian Correspondence of the Relativistic Analyses} 
                                    \label{sec:Six-gauges}

In \cite{MDE,Hyun} we developed arguments on the correspondences
between the Newtonian and the relativistic analyses.
In order to reinforce the Newtonian correspondences of certain
(gauge-invariant) variables in certain gauges, in the following we
will present the closed form differential equations for 
$\delta$, $v$ and $\varphi$ in the six different gauge conditions
and compare them with the Newtonian ones in 
Eqs. (\ref{Newt-delta-eq}-\ref{Newt-Phi-eq}).

There are reasons why we should know about possible Newtonian behaviors
of the relativistic variables.
Variables in the relativitistic gravity are only parameters appearing in
the metric and energy-momentum tensor.
{}For example, later we will find that only in the zero-shear gauge does
the perturbed curvature variable behave as the perturbed Newtonian potential,
which we have some experience of in Newtonian physics.
The same variable evaluated in other gauge conditions is simply other
variable and we cannot regard it as the perturbed potential.
As we have introduced several different fundamental gauge conditions
it became necessary to identify the correct gauge where a variable shows
the corresponding Newtonian behavior.
As will be shown in this section, we rarely find Newtonian correspondences.
In fact, there exists an almost unique gauge condition for each
relativistic perturbation variable which shows Newtonian behavior.
The results will be summarized in Sec. \ref{sec:Correspondences}.

In this section we consider a {\it pressureless medium},
$p = 0$ (thus ${\rm w} = 0 = c_s^2$) and $e = 0 = \sigma$.
However, we consider general $K$ and $\Lambda$.

\subsection{Comoving Gauge} 
                                    \label{sec:CG}

As the gauge condition we set $v/k \equiv 0$.
Equivalently, we can set $v/k = 0$ and leave the other variables as
the gauge-invariant combinations with subindices $v/k$ or simply $v$.
{}From Eq. (\ref{eq7}) we have $\alpha_v = 0$.
Thus the comoving gauge is a case of the synchronous gauge; this is
true {\it only} in a pressureless situation.
{}From Eqs. (\ref{eq1}-\ref{eq7}) we can derive:
\bea
   & & \ddot \delta_v + 2 H \dot \delta_v - 4 \pi G \mu \delta_v = 0,
   \label{CG-delta-eq} \\
   & & \ddot \varphi_v + 3 H \dot \varphi_v - {K \over a^2} \varphi_v = 0.
   \label{CG-varphi-eq} 
\eea
Thus, Eq. (\ref{CG-delta-eq}) has a form identical to Eq. 
(\ref{Newt-delta-eq}).
However, Eq. (\ref{CG-varphi-eq}) differs from Eq. (\ref{Newt-Phi-eq}),
and apparently we do not have an equation for $v$.

{}For $K = 0$ Eq. (\ref{CG-varphi-eq}) leads to
two solutions which are $\varphi_v \propto {\rm constant}$ and 
$\int_0^t a^{-3} dt$. 
Since we have $\alpha_v = 0$, for $K = 0$, from Eqs. (\ref{eq1},\ref{eq3})
we have $\dot \varphi_v = 0$.
This implies that, for $K = 0$, the second solution, 
$\varphi_v \propto \int_0^t a^{-3} dt$, should have the vanishing coefficient,
see Eq. (\ref{varphi_v-sol3}).
The solution of Eq. (\ref{CG-varphi-eq}) for general $K$ will be
presented later in Eq. (\ref{varphi_v-sol}).

\subsection{Synchronous Gauge} 
                                    \label{sec:SG}

We let $\alpha \equiv 0$.
This gauge condition does not fix the temporal gauge transformation property
completely.
Thus, although we can still indicate the variables in this gauge condition
using subindices $\alpha$ without ambiguity, these variables are not 
gauge-invariant; see the discussion after Eq. (\ref{varphi-GI}).
Equation (\ref{eq7}) leads to
\bea
   v_\alpha = c_g {k \over a},
\eea
which is a pure gauge mode.
Thus, fixing $c_g \equiv 0$ exactly corresponds to the comoving gauge.
We can show that the following two equations are not affected by the
remaining gauge mode.
\bea
   & & \ddot \delta_\alpha + 2 H \dot \delta_\alpha 
       - 4 \pi G \mu \delta_\alpha = 0,
   \label{SG-delta-eq} \\
   & & \ddot \varphi_\alpha + 3 H \dot \varphi_\alpha 
       - {K \over a^2} \varphi_\alpha = 0.
   \label{SG-varphi-eq} 
\eea
Equation (\ref{SG-delta-eq}) is identical to Eq. (\ref{Newt-delta-eq}).
This is because in a pressureless medium the behavior of the gauge modes 
happen to coincide with the behavior of the decaying physical solutions
for $\delta_\alpha$ and $\varphi_\alpha$.
However, for the variables $\kappa_\alpha$ and $\chi_\alpha$ the gauge mode 
contribution appears explicitly.

Thus, in a pressureless medium, variables
in the synchronous gauge behave the same as the ones in the comoving gauge, 
except for the additional gauge modes which appear in the synchronous gauge
for some variables.
In a pressureless medium, we can simultaneously impose both the comoving gauge
and the synchronous gauge conditions.
However, this is possible only in a pressureless medium, see the Appendix.
In Sec. \ref{sec:Discussion} we indicate some common errors in the literature
based on the synchronous gauge.

\subsection{Zero-shear Gauge} 
                                    \label{sec:ZSG}

We let $\chi \equiv 0$, and substitute the other variables into the 
gauge-invariant combinations with subindices $\chi$.
We can derive:
\bea
   & & \ddot \delta_\chi + { 2 k^2 / a^2 - 36 \pi G \mu \left[ 1 + ( H^2 
       - 2 \dot H ) a^2 / \left( k^2 - 3 K \right) \right] \over
       k^2 / a^2 - 12 \pi G \mu \left[ 1 + 3 H^2 a^2/\left( k^2 - 3K \right) 
       \right] }
       H \left( \dot \delta_\chi 
       - 12 \pi G \mu {a^2 \over k^2 - 3 K} H \delta_\chi \right)
   \nonumber \\
   & & \qquad 
       - 4 \pi G \mu \left[ 1 - 3 ( H^2 - 2 \dot H )
       {a^2 \over k^2 - 3 K} \right] \delta_\chi = 0,
   \label{ZSG-delta-eq} \\
   & & \ddot v_\chi + 3 H \dot v_\chi 
       + \left( \dot H + 2 H^2 - 4 \pi G \mu \right) v_\chi = 0,
   \label{ZSG-v-eq} \\
   & & \ddot \varphi_\chi + 4 H \dot \varphi_\chi + \left( \dot H + 3 H^2 
       - 4 \pi G \mu \right) \varphi_\chi = 0.
   \label{ZSG-varphi-eq} 
\eea
Thus, Eqs. (\ref{ZSG-v-eq},\ref{ZSG-varphi-eq}) have forms identical 
to Eqs. (\ref{Newt-v-eq},\ref{Newt-Phi-eq}), respectively.
However, only in the small-scale limit is the behavior of $\delta_\chi$
the same as the Newtonian one.

\subsection{Uniform-expansion Gauge} 
                                    \label{sec:UEG}

We let $\kappa \equiv 0$, and substitute the remaining variables into
the gauge-invariant combinations with subindices $\kappa$.
We can derive:
\bea
   & & \ddot \delta_\kappa 
       + \left( 2 H - {12 \pi G \mu H \over k^2/a^2 - 3 \dot H}
       \right) \dot \delta_\kappa - \left[ 4 \pi G \mu {k^2/a^2 + 6 H^2 \over
       k^2/a^2 - 3 \dot H} + \left( {12 \pi G \mu H \over
       k^2/a^2 - 3 \dot H} \right)^\cdot \right] \delta_\kappa = 0,
   \label{UEG-delta-eq} \\ 
   & & \ddot v_\kappa + \left[ 2 H - {12 \pi G \mu H \over k^2/a^2 - 3 \dot H}
       + {4 \pi G \mu \over k^2/a^2 - 3 \dot H}
       \left( {k^2/a^2 - 3 \dot H \over 4 \pi G \mu} \right)^\cdot
       \right] \dot v_\kappa
   \nonumber \\ 
   & & \quad 
       + \left[ \dot H + H^2 - 4 \pi G \mu {k^2/a^2 + 3 H^2 \over
       k^2/a^2 - 3 \dot H} + H {4 \pi G \mu \over k^2/a^2 - 3 \dot H}
       \left( { k^2/a^2 - 3 \dot H \over 4 \pi G \mu} \right)^\cdot
       \right] v_\kappa = 0,
   \label{UEG-v-eq} \\ 
   & & \ddot \varphi_\kappa 
       + \left( 4 H - {12 \pi G \mu H \over k^2/a^2 - 3 \dot H}
       \right) \dot \varphi_\kappa
       + \left[ \dot H + 3 H^2 - 4 \pi G \mu {k^2/a^2 + 9 H^2 \over
       k^2/a^2 - 3 \dot H} - \left( {12 \pi G \mu H \over k^2/a^2
       - 3 \dot H} \right)^\cdot \right] \varphi_\kappa = 0.
   \label{UEG-varphi-eq}
\eea
In the small-scale limit we can show that
Eqs. (\ref{UEG-delta-eq}-\ref{UEG-varphi-eq}) 
reduce to Eqs. (\ref{Newt-delta-eq}-\ref{Newt-Phi-eq}), respectively.
Thus, {\it in the small-scale limit}, all three variables $\delta_\kappa$, 
$v_\kappa$ and $\varphi_\kappa$ correctly reproduce the Newtonian behavior.
However, outside or near the (visual) horizon scale, the behaviors of all these
variables {\it strongly deviate} from the Newtonian ones.

In Sec. 84 of \cite{Peebles-1980} we find that in order to get the usual 
Newtonian equations a coordinate transformation is made so that
we have $\dot h \equiv 0$ in the new coordinate.
We can show that $\dot h = 2 \kappa$ in our notation.
Thus, the new coordinate used in \cite{Peebles-1980} is the uniform-expansion 
gauge\footnote{This was incorrectly pointed out after Eq. (49) in \cite{MDE}:
         In \cite{MDE} it was wishfully mentioned that in Sec. 84 of 
         \cite{Peebles-1980} the gauge transformations were made into the 
         comoving gauge for $\delta$ and into the zero-shear gauge for 
         $v$ and $\varphi$, respectively.
     }.

\subsection{Uniform-curvature Gauge} 
                                    \label{sec:UCG}

We let $\varphi \equiv 0$, and substitute the other variables into the 
gauge-invariant combinations with subindices $\varphi$.
We have
\bea
   & & \ddot \delta_\varphi + 2 H { (k^2 - 3K)/a^2 + 18 \pi G \mu \over 
       (k^2 - 3K)/a^2 + 12 \pi G \mu } \dot \delta_\varphi
       - { 4 \pi G \mu k^2/a^2 \over 
       (k^2 - 3K)/a^2 + 12 \pi G \mu } \delta_\varphi = 0,
   \label{UCG-delta-eq} \\
   & & \ddot v_\varphi + \left( 5 H + 2 {\dot H \over H} \right) \dot v_\varphi
       + \left( 3 \dot H + 4 H^2 \right) v_\varphi = 0. 
   \label{UCG-v-eq}
\eea
In the small-scale limit Eq. (\ref{UCG-delta-eq}) reduces to
Eq. (\ref{Newt-delta-eq}).
In the uniform-curvature gauge, the perturbed potential is set to be equal
to zero by the gauge condition.
The uniform-curvature gauge condition is known to have 
distinct properties in handling the scalar field or the dilaton field 
which appears in a broad class of the generalized gravity theories
\cite{MSF-GGT,GRG-MSF-GGT}.

\subsection{Uniform-density Gauge} 
                                    \label{sec:UDG}

We let $\delta \equiv 0$, and substitute the other variables into the 
gauge-invariant combinations with subindices $\delta$.
We have
\bea
   & & \ddot v_\delta + 2 \left( 2 H + {\dot H \over H} \right) \dot v_\delta
       + 3 \left( H^2 + \dot H \right) v_\delta = 0,
   \label{UDG-v-eq} \\
   & & \ddot \varphi_\delta + 2 H { (k^2 - 3 K)/a^2 + 18 \pi G \mu
       \over (k^2- 3K)/a^2 + 12 \pi G \mu} \dot \varphi_\delta
       - {4 \pi G \mu k^2/a^2
       \over (k^2- 3K)/a^2 + 12 \pi G \mu} \varphi_\delta = 0.
   \label{UDG-varphi-eq} 
\eea
These equations differ from Eqs. (\ref{Newt-v-eq},\ref{Newt-Phi-eq}).
Of course, we have no equation for $\delta$ which is set to be equal to zero 
by our choice of the gauge condition.

\subsection{Newtonian Correspondences} 
                                    \label{sec:Correspondences}

After a thorough comparison made in this section we found that 
equations for $\delta$ in the comoving gauge ($\delta_v$), and
for $v$ and $\varphi$ in the zero-shear gauge ($v_\chi$ and $\varphi_\chi$)
show the same forms as the corresponding Newtonian equations 
{\it in general scales} and considering general $K$ and $\Lambda$.
Using these gauge-invariant combinations in Eq. (\ref{GI-combinations}),
Eqs. (\ref{eq1}-\ref{eq7}) can be combined to give:
\bea
   \dot \delta_v = - {k^2 - 3K \over a k} v_\chi, \quad
       \dot v_\chi + H v_\chi = - {k \over a} \varphi_\chi, \quad
       {k^2 - 3K \over a^2} \varphi_\chi = 4 \pi G \mu \delta_v.
   \label{GI-eqs}
\eea
Comparing Eq. (\ref{GI-eqs}) with Eq. (\ref{Newt-eqs}) 
we can identify either one of the following correspondences:
\bea
   & & \delta_v \leftrightarrow \delta |_{\rm Newtonian}, \quad
       {k^2 - 3 K \over k^2} v_\chi 
       \leftrightarrow \delta v |_{\rm Newtonian}, \quad
       - {k^2 - 3 K \over k^2} \varphi_\chi \leftrightarrow 
       \delta \Phi |_{\rm Newtonian},
   \label{Correspondences-1} \\
   & & {k^2 \over k^2 - 3 K} \delta_v \leftrightarrow 
       \delta |_{\rm Newtonian}, \quad
       v_\chi \leftrightarrow \delta v |_{\rm Newtonian}, \quad
       - \varphi_\chi \leftrightarrow \delta \Phi |_{\rm Newtonian}.
   \label{Correspondences-2}
\eea
In \cite{MDE} we proposed the correspondences in Eq. (\ref{Correspondences-1}),
but the ones in Eq. (\ref{Correspondences-2}) also work well.
In fact, the gravitational potential identified in 
Eq. (\ref{Correspondences-2}) is the one often found in the Newtonian limit 
of the general relativity, e.g. in \cite{Weinberg}.
Using Eqs. (\ref{GI-combinations},\ref{eq3},\ref{eq4},\ref{GI-eqs}) 
we can also identify the following relations:
\bea
   & & \delta_v = 3 H {a \over k} v_\delta, \quad
       v_\chi = - {k \over a} \chi_v = - {ak \over k^2 - 3 K} \kappa_v, \quad
       \varphi_\chi = -\alpha_\chi = - H \chi_\varphi.
\eea

{}From a given solution known in one gauge we can derive all the rest of the 
solutions even in other gauge conditions.
This can be done either by using the gauge-invariant combination of variables 
or directly through gauge transformations.
The complete set of exact solutions in a pressureless medium is presented 
in \cite{MDE}.
{}From our study in this section and using the complete solutions presented 
in Tables 1 and 2 of \cite{MDE} we can identify variables in certain gauges 
which correspond to the Newtonian ones.  These are summarized in Table 1.

\centerline{ [[{\bf TABLE 1.}]] }

\noindent
As mentioned earlier, all three variables in the uniform-expansion gauge show 
Newtonian correspondence in the small-scale; however all of them change 
the behaviors near and above the horizon scale. 
In the small-scale limit, except for the  
uniform-density gauge where $\delta \equiv 0$, 
$\delta$ in all gauge conditions behaves in the same way \cite{Bardeen-1980}.
Meanwhile, since only $\delta_v$ ($\delta$ in the comoving gauge)
shows the Newtonian behavior in a general scale, it may be natural to identify 
$\delta_v$ as the one most closely resembling the Newtonian one.

Notice that, although we have horizons in the relativistic analysis
the equations for $\delta_v$, $v_\chi$ and $\varphi_\chi$
keep the same form as the corresponding Newtonian equations
{\it in the general scale}.
Considering this as an additional point we regard
$\delta_v$, $v_\chi$ and $\varphi_\chi$ as the one most closely corresponding
to the Newtonian variables.
This argument will become stronger as we consider the case with a general
pressure in the next section.

\section{Relativistic Cosmological Hydrodynamics}
                                \label{sec:Hydrodynamics}

\subsection{General Equations and Large-scale Solutions}
                                \label{sec:Eqs}

In the previous sections we have shown that the gauge-invariant
combinations $\delta_v$, $v_\chi$ and $\varphi_\chi$ behave most similarly
to the Newtonian $\delta \equiv \delta \varrho/\varrho$, 
$\delta v$ and $\delta \Phi$, respectively.
The equations remain the same in a general scale which includes the
superhorizon scales in the relativistic situation considering 
general $K$ and $\Lambda$.
In this section, we will present the fully general relativistic equations 
for $\delta_v$, $v_\chi$ and $\varphi_\chi$ including the effects of the 
general pressure terms.
Equations (\ref{eq1}-\ref{eq7}) are the basic set of 
perturbation equations in a gauge-ready form.

{}From Eqs. (\ref{eq3},\ref{eq6},\ref{eq7}),
Eqs. (\ref{eq4},\ref{eq7}), Eqs. (\ref{eq2},\ref{eq3}), and
Eqs. (\ref{eq1},\ref{eq3},\ref{eq4}), we have, respectively:
\bea
   & & \dot \delta_v - 3 H {\rm w} \delta_v
       = - \left( 1 + {\rm w} \right) {k \over a} {k^2 - 3 K \over k^2} v_\chi
       - 2 H {k^2 - 3K \over a^2} {\sigma \over \mu},
   \label{P-1} \\
   & & \dot v_\chi + H v_\chi = - {k \over a} \varphi_\chi
       + {k \over a \left( 1 + {\rm w} \right)} 
       \left[ c_s^2 \delta_v + {e \over \mu}
       - 8 \pi G \left( 1 + {\rm w} \right) \sigma
       - {2 \over 3} {k^2 - 3 K \over a^2} {\sigma \over \mu} \right],
   \label{P-2} \\
   & & {k^2 - 3 K \over a^2} \varphi_\chi = 4 \pi G \mu \delta_v,
   \label{P-3} \\
   & & \dot \varphi_\chi + H \varphi_\chi
       = - 4 \pi G \left( \mu + p \right) {a \over k} v_\chi 
       - 8 \pi G H \sigma.
   \label{P-4}
\eea
Considering either one of the correspondences in 
Eqs. (\ref{Correspondences-1},\ref{Correspondences-2})
we immediately see that Eqs. (\ref{P-1}-\ref{P-4}) have the 
correct Newtonian limit expressed in Eqs. (\ref{Newt-eqs},\ref{GI-eqs}).
Equations (\ref{P-1}-\ref{P-4}) were presented in \cite{Bardeen-1980}.

Combining Eqs. (\ref{P-1}-\ref{P-4}) we can derive closed
form expressions for the $\delta_v$ and $\varphi_\chi$
which are the relativistic counterpart of Eqs. 
(\ref{Newt-delta-eq},\ref{Newt-Phi-eq}).
We have:
\bea
   & & \ddot \delta_v + \left( 2 + 3 c_s^2 - 6 {\rm w} \right) H \dot \delta_v
       + \Bigg[ c_s^2 {k^2 \over a^2} 
       - 4 \pi G \mu \left( 1 - 6 c_s^2 + 8 {\rm w} - 3 {\rm w}^2 \right)
       + 12 \left( {\rm w} - c_s^2 \right) {K \over a^2}
       + \left( 3 c_s^2 - 5 {\rm w} \right) \Lambda \Bigg] \delta_v
   \nonumber \\
   & & \qquad \qquad
       = {1 + {\rm w} \over a^2 H} \left[ {H^2 \over a (\mu + p)} 
       \left( {a^3 \mu \over H} \delta_v \right)^\cdot \right]^\cdot
       + c_s^2 {k^2 \over a^2} \delta_v
   \nonumber \\
   & & \qquad \qquad
       = - {k^2 - 3 K \over a^2} { 1 \over \mu }
       \left\{ e + 2 H \dot \sigma
       + 2 \left[ - {1 \over 3} {k^2 \over a^2} + 2 \dot H
       + 3 \left( 1 + c_s^2 \right) H^2 \right] \sigma \right\},
   \label{delta-eq}\\
   & & \ddot \varphi_\chi + \left( 4 + 3 c_s^2 \right) H \dot \varphi_\chi
       + \left[ c_s^2 {k^2 \over a^2} 
       + 8 \pi G \mu \left( c_s^2 - {\rm w} \right)
       - 2 \left( 1 + 3 c_s^2 \right) {K \over a^2}
       + \left( 1 + c_s^2 \right) \Lambda \right] \varphi_\chi
   \nonumber \\
   & & \qquad \qquad
       = {\mu + p \over H} \left[ {H^2 \over a (\mu + p)} 
       \left( {a \over H} \varphi_\chi \right)^\cdot \right]^\cdot
       + c_s^2 {k^2 \over a^2} \varphi_\chi
       = {\rm stresses}.
   \label{varphi-eq}
\eea
These two equations are related by Eq. (\ref{P-3}) which resembles the 
Poisson equation.
Notice the remarkably compact forms presented in the second steps of
Eqs. (\ref{delta-eq},\ref{varphi-eq}).
It may be an interesting exercise to show that the above equations
are indeed valid for general $K$ and $\Lambda$, and
for the general equation of state $p = p(\mu)$;
use $\dot {\rm w} = - 3 H (1 + {\rm w}) ( c_s^2 - {\rm w} )$.
Equation (\ref{delta-eq}) became widely known through Bardeen's seminal paper 
in \cite{Bardeen-1980}.
In a less general context but originally 
Eqs. (\ref{delta-eq},\ref{varphi-eq}) were derived by
Nariai \cite{Nariai} and Harrison \cite{Harrison}, respectively;
however, the compact expressions are new results in this paper.

If we ignore the entropic and anisotropic pressures ($e = 0 = \sigma$) 
on scales larger than Jeans scale
Eq. (\ref{varphi-eq}) immediately leads to a general integral form 
solution as
\bea
   \varphi_\chi ({\bf k}, t)
   &=& 4 \pi G C ({\bf k}) {H \over a} \int_0^t {a (\mu + p) \over H^2} dt
       + {H \over a} d ({\bf k})
   \nonumber \\
   &=& C ({\bf k}) \left[ 1 - {H \over a} \int_0^t
       a \left( 1 - {K \over \dot a^2} \right) dt \right]
       + {H \over a} d ({\bf k}).
   \label{varphi_chi-sol}
\eea
This solution was first derived in Eq. (108) of 
\cite{HV}\footnote{We correct a typographical 
         error in Eq. (129) of \cite{HV}: $c_0/a$ should be replaced by $c_0$.
         },
see also Eq. (55) in \cite{PRW}.
Solutions for $\delta_v$ and $v_\chi$ follow from Eqs.
(\ref{P-3},\ref{P-4}), respectively, as
\bea
   & & \delta_v ({\bf k}, t) = {k^2 - 3 K \over 4 \pi G \mu a^2} 
       \varphi_\chi ({\bf k}, t), 
   \label{delta_v-sol} \\
   & & v_\chi ({\bf k}, t)= - {k \over 4 \pi G ( \mu + p) a^2}
       \left\{ C ({\bf k}) \left[ {K \over \dot a} 
       - \dot H \int_0^t a \left( 1 - {K \over \dot a^2} \right) dt \right]
       + \dot H d ({\bf k}) \right\}.
   \label{v_chi-sol}
\eea
We stress that these large-scale asymptotic solutions are
{\it valid for the general $K$, $\Lambda$, and $p = p(\mu)$};
$C({\bf k})$ and $d({\bf k})$ are integration constants considering 
the general evolution of $p = p(\mu)$.
We also emphasize that {\it the large-scale criterion is the sound-horizon},
and thus in the matter dominated era the solutions are valid even far 
inside the (visual) horizon as long as the scales are larger than Jeans scale.
In the pressureless limit, considering Eq. (\ref{Correspondences-2}),
Eqs. (\ref{varphi_chi-sol}-\ref{v_chi-sol})
correctly reproduce Eqs. (\ref{delta-N}-\ref{Phi-N}).

Using the set of gauge-ready form equations in
Eqs. (\ref{eq1}-\ref{eq7}) the complete set of corresponding general
solutions for all variables in all different gauges can be easily derived;
for such sets of solutions with less general assumptions see \cite{MDE,Ideal}.
{}For $K = 0 = \Lambda$ and ${\rm w} = {\rm constant}$ we have
$a \propto t^{2/[3(1+{\rm w})]}$ and Eqs. 
(\ref{varphi_chi-sol}-\ref{v_chi-sol}) become:
\bea
   \varphi_\chi ({\bf k}, t) 
   &=& {3 (1 + {\rm w}) \over 5 + 3 {\rm w}} C ({\bf k})
       + {2 \over 3 (1 + {\rm w})} {1 \over at} d ({\bf k})
   \nonumber \\
   &\propto& {\rm constant}, 
       \quad t^{- {5 + 3 {\rm w} \over 3 ( 1 + {\rm w} )} }
       \quad \propto \quad {\rm constant}, 
       \quad a^{- {5 + 3 {\rm w} \over 2} },
   \nonumber \\
   \delta_v ({\bf k}, t) 
   &\propto& t^{2(1 + 3 {\rm w}) \over 3 (1 + {\rm w})}, 
       \quad t^{- {1 - {\rm w} \over 1 + {\rm w} } }
       \quad \propto \quad a^{1 + 3 {\rm w}}, 
       \quad a^{- {3 \over 2} (1 - {\rm w}) },
   \nonumber \\
   v_\chi ({\bf k}, t)
   &\propto& t^{1 + 3 {\rm w} \over 3 (1 + {\rm w})}, 
       \quad t^{- {4 \over 3(1 + {\rm w}) } }
       \quad \propto \quad a^{1 + 3 {\rm w} \over 2}, 
       \quad a^{- 2}.
   \label{Sol-w=const}
\eea
These solutions were presented in \cite{Bardeen-1980} and
in less general contexts but originally in \cite{Harrison,Nariai}.
{}For ${\rm w} =0$ we recover the well known Newtonian behaviors.

\subsection{Conserved Quantities}
                                        \label{sec:Conservation}

In an ideal fluid, the curvature fluctuations in several gauge conditions 
are known to be conserved in the {\it superhorizon scale} independently
of the changes in the background equation of state.
{}From Eqs. (41,73) in \cite{Ideal} we find [for $K = 0 = \Lambda$,
but for general $p(\mu)$]:
\bea
   \varphi_v = \varphi_\delta = \varphi_\kappa = \varphi_\alpha = C ({\bf k}),
   \label{varphi-conserv}
\eea
and the dominating decaying solutions vanish (or, are cancelled).
The combinations follow from Eq. (\ref{GT}) as:
\bea
   \varphi_v = \varphi - {aH \over k} v, \quad
       \varphi_\delta \equiv \varphi +{\delta \over 3 (1 + {\rm w})} 
       \equiv \zeta, \quad
       \varphi_\kappa \equiv \varphi + {H \over 3 \dot H - k^2/a^2 } \kappa, 
       \quad
       \varphi_\alpha \equiv \varphi - H \int^t \alpha dt.
   \label{varphi-GI}
\eea
The $\varphi_\alpha$ combination which is $\varphi$ in the synchronous gauge
is not gauge-invariant; the lower bound of integration gives the behavior
of the gauge mode; in Eq. (\ref{varphi-conserv}) we ignored the gauge mode.
The large-scale conserved quantity, $\zeta$, proposed in
\cite{Bardeen-1988,BST} is $\varphi_\delta$.
In \cite{Ideal} the above conservation properties are shown assuming 
$K = 0 = \Lambda$; in such a case the integral form solutions for a complete
set of variables are presented in Table 8 in \cite{Ideal}.
In Eqs. (\ref{varphi_chi-sol}-\ref{v_chi-sol}) we have the large-scale 
integral form solutions in the case of general $K$ and $\Lambda$.
Thus, now we can see the behavior of these variables considering the 
general $K$ and $\Lambda$.

{}Evaluating $\varphi_v$ in Eq. (\ref{varphi-GI}) in the zero-shear gauge,
thus $\varphi_v = \varphi_\chi - (aH / k) v_\chi$, and using
the solutions in Eqs. (\ref{varphi_chi-sol},\ref{v_chi-sol}) we 
can derive
\bea
   \varphi_v ({\bf k}, t) 
   &=& C ({\bf k}) \left\{ 1 + {K \over a^2}
       {1 \over 4 \pi G (\mu + p)} \left[ 1 
       - {H \over a} \int_0^t a \left( 1 - {K \over \dot a^2} \right) dt 
       \right] \right\}
       + {K \over a^2} { H/a \over 4 \pi G ( \mu + p)} d ({\bf k})
   \nonumber \\
   &=& C ({\bf k}) \left[ 1 + {K \over a^2}
       {H/a \over \mu + p} \int_0^t {a (\mu + p) \over H^2} dt \right]
       + {K \over a^2} { H/a \over 4 \pi G ( \mu + p)} d ({\bf k}).
   \label{varphi_v-sol}
\eea
Thus, for $K = 0$ (but for general $\Lambda$) we have
\bea
   \varphi_v ({\bf k}, t) = C ({\bf k}),
   \label{varphi_v-sol2}
\eea
with the vanishing decaying solution. 
The disappearance of the decaying 
solution in Eq. (\ref{varphi_v-sol2}) implies that the dominating decaying 
solutions in Eqs. (\ref{varphi_chi-sol},\ref{v_chi-sol}) cancel out 
for $K = 0$.
In fact, for $K = 0$, from Eqs. 
(\ref{eq1},\ref{eq2},\ref{eq3},\ref{eq6},\ref{eq7}) we can derive
\bea
    & & {c_s^2 H^2 \over a^3 ( \mu + p)} \left[ {a^3 ( \mu + p) \over c_s^2 H^2}
        \dot \varphi_v \right]^\cdot
        + c_s^2 {k^2 \over a^2} \varphi_v = {\rm stresses}.
    \label{varphi_v-fluid-eq}
\eea
{}For a pressureless case ($c_s^2 = 0$), instead of
Eq. (\ref{varphi_v-fluid-eq}) we have $\dot \varphi_v = 0$;
see after Eq. (\ref{CG-varphi-eq}).
In the large-scale limit, and ignoring the stresses, we have an integral
form solution
\bea
   & & \varphi_v
       = C ({\bf x}) - \tilde D ({\bf x})
       \int^t_0 {c_s^2 H^2 \over a^3 ( \mu + p ) } dt.
   \label{varphi_v-sol3}
\eea
Compared with the solution in Eq. (\ref{varphi_v-sol2}), the $\tilde D$ 
term in Eq. (\ref{varphi_v-sol3}) is $c_s^2(k/aH)^2$ order higher
than $d$ terms in the other gauge.
Therefore, for $K = 0$, $\varphi_v$ is conserved for the generally varying
background equation of state, i.e., for general $p = p(\mu)$.
Since the solutions in 
Eqs. (\ref{varphi_chi-sol},\ref{delta_v-sol},\ref{v_chi-sol}) are valid 
in super-sound-horizon scale, the conservation property in 
Eq. (\ref{varphi_v-sol2}) is valid in all scales in the matter 
dominated era.

{}For $\varphi_\delta$, by evaluating it in the comoving gauge we have
\bea
   \varphi_\delta \equiv \varphi_v + {\delta_v \over 3 (1 + {\rm w})}.
\eea
Using Eqs. (\ref{varphi_v-sol},\ref{delta_v-sol}) we have
\bea
   \varphi_\delta ({\bf k}, t)
   &=& C ({\bf k}) \left\{ 1 + {k^2 \over a^2}
       {1 \over 12 \pi G (\mu + p)} \left[ 1
       - {H \over a} \int_0^t a \left( 1 - {K \over \dot a^2} \right) dt
       \right] \right\}
       + {k^2 \over a^2} { H/a \over 12 \pi G ( \mu + p)} d ({\bf k}).
   \label{varphi_delta-sol}
\eea
Since the higher order term ignored in Eq. (\ref{varphi_v-sol}) 
is $c_s^2 (k/aH)^2$ order higher, in the medium with $c_s \sim 1 (\equiv c)$,
the terms involving $(k/a)^2/(\mu + p)$ are not necessarily valid.
Thus, to the leading order in the superhorizon scale we have
\bea
   \varphi_\delta ({\bf k}, t) = C ({\bf k}).
\eea
Thus, in the superhorizon scale $\varphi_\delta$ is conserved 
apparently  considering $K$ and $\Lambda$.
[In the case of nonvanishing $K$ we may need to be 
       careful in defining the superhorizon scale. 
       Here, as the superhorizon condition we simply {\it took} $k^2$ term 
       to be negligible.
       In a hyperbolic (negatively curved) space with $K = -1$, 
       since $R^{(3)} = {6 K / a^2}$, the distance 
       $a/\sqrt{|K|}$ introduces a curvature scale: on smaller scales the 
       space is near flat and on larger scales the curvature term dominates.
       In the unit of the curvature scale $k^2$ less than one ($\sim |K|$)
       corresponds to the super-curvature scale, and $k^2$ larger than 
       one corresponds to the sub-curvature scale.  
       Thus, when we ignored the $k^2$ term compared with $|K|$ in 
       a negatively curved space, we are considering the super-curvature scale.
       In other words, for the hyperbolic backgrounds $k^2$ takes continuous 
       value with $k^2 \ge 0$ where $k^2 \ge 1$ corresponds to subcurvature 
       mode whereas $0\le k^2 <1 $ corresponds to supercurvature mode, and 
       our large-scale limit took $k^2 \rightarrow 0$.
       A useful study in the hyperbolic situation can be found 
       in \cite{Lyth-etal}.]
However, in a medium with the negligible sound speed (like the matter dominated 
era), Eq. (\ref{varphi_delta-sol}) is valid considering the $k^2$ terms.
Thus, as the scale enters the horizon in the matter dominated era 
$\varphi_\delta$ wildly changes its behavior and is {\it no longer conserved 
inside the horizon};
inside the horizon $\varphi_\delta$ is dominated by the $\delta$ part
which shows Newtonian behavior of $\delta$ in most of the gauge conditions, 
see Table 1.

{}For $\varphi_\kappa$, by evaluating it in the uniform-density gauge
and using Eq. (\ref{eq2}), which gives 
$\kappa_\delta = (k^2 - 3K)/(a^2 H) \varphi_\delta$, we have
\bea
   \varphi_\kappa = \varphi_\delta
       \left[ 1 + {k^2 - 3K \over 12 \pi G (\mu + p) a^2} \right]^{-1}.
   \label{varphi_kappa-sol}
\eea
Thus, in the superhorizon scale and for $K = 0$, $\varphi_\kappa$ is conserved.
As in $\varphi_\delta$, in a medium with the negligible sound speed, 
the solution in Eq. (\ref{varphi_kappa-sol}) with Eq. (\ref{varphi_delta-sol}) 
is valid considering the $k^2$ terms.
Thus, as the scale enters the horizon in the matter dominated era 
$\varphi_\kappa$
also changes its behavior and is {\it no longer conserved inside the horizon}.

Now we summarize:
In the superhorizon scale $\varphi_v$ and $\varphi_\kappa$ are conserved for 
$K = 0$, while $\varphi_\delta$ is conserved considering the general $K$.
In the matter dominated era, for $K = 0$, $\varphi_v$ is still conserved in the
super-sound-horizon scale which virtually covers all scales,
whereas $\varphi_\delta$ and $\varphi_\kappa$ change their behaviors
near the horizon crossing epoch and are no longer conserved inside the horizon.
In this regard, we may say, $\varphi_v$ is the {\it best conserved quantity}.
Curiously, although $\varphi_\chi$ is the variable most closely resembling
the Newtonian gravitational potential, it is not conserved for changing 
equation of state, see Eq. (\ref{varphi_chi-sol});
however, it is conserved independently of the horizon crossing in the
matter dominated era for $K = 0 = \Lambda$, see Eq. (\ref{Sol-w=const}).

Similar conservation properties of the curvature variable in certain gauge
conditions remain true for models based on a minimally coupled scalar 
field and even on classes of generalized gravity theories.
In the generalized gravity the uniform-field gauge is more suitable
for handling the conservation property, and the uniform-field gauge
coincides with the comoving gauge in the minimally coupled scalar field.
{}Thorough studies of the minimally coupled scalar field
and the generalized gravity, assuming $K = 0 = \Lambda$, are made in 
\cite{MSF-GGT,MSF-GGT-2} and summarized in \cite{GRG-MSF-GGT}.

\subsection{Multi-component situation}
                               \label{sec:Multi}

We consider the energy-momentum tensor composed of multiple components as
\bea
   T_{ab} = \sum_{(l)} T_{(l)ab}, \quad
       T^{\;\;\;\; b}_{(i)a;b} \equiv Q_{(i)a}, \quad
       \sum_{(l)} Q_{(l)a} = 0,
   \label{Tab-multi}
\eea
where $i,l (= 1,2, \dots, n)$ indicates $n$ fluid components, and
$Q_{(i)a}$ considers possible mutual interactions among fluids.
Since we are considering the scalar-type perturbation we decompose 
the $Q_{(i)a}$ in the following way:
\bea
   Q_{(i)0} \equiv - a (1 + \alpha) Q_{(i)}, \quad
       Q_{(i)} \equiv \bar Q_{(i)} + \delta Q_{(i)}, \quad
       Q_{(i)\alpha} \equiv J_{(i),\alpha}.
\eea
The scalar-type fluid quantities in Eq. (\ref{Tab}) can be considered 
as the collective ones which are related to the individual component as:
\bea
   & & \bar \mu = \sum_{(l)} \bar \mu_{(l)}, \quad
       \bar p = \sum_{(l)} \bar p_{(l)}; \quad
       \delta \mu = \sum_{(l)} \delta \mu_{(l)}, \quad
       \delta p = \sum_{(l)} \delta p_{(l)}, 
   \nonumber \\
   & & (\mu + p) v = \sum_{(l)} (\mu_{(l)} + p_{(l)} ) v_{(l)}, \quad
       \sigma = \sum_{(l)} \sigma_{(l)}.
   \label{fluid-multi}
\eea

{}For the background Eq. (\ref{BG-eqs}) remains valid for the collective fluid
quantities.
An additional equation follows from the individual
energy-momentum conservation in Eq. (\ref{Tab-multi}) as
\bea
   \dot \mu_{(i)} + 3 H \left( \mu_{(i)} + p_{(i)} \right) = Q_{(i)}.
   \label{BG-multi}
\eea
{}For the perturbations Eqs. (\ref{eq1}-\ref{eq7}) remain valid for the 
collective fluid quantities. 
The additional equations we need follow from the individual
energy-momentum conservation in Eq. (\ref{Tab-multi}).
{}From $T^{\;\;\;\; b}_{(i)0;b} = Q_{(i)0}$ and using Eq. (\ref{eq1})
we have the energy conservation of the fluid components 
\bea
   \delta \dot \mu_{(i)} + 3 H \left( \delta \mu_{(i)}
       + \delta p_{(i)} \right)
       = - {k \over a} \left( \mu_{(i)} + p_{(i)} \right) v_{(i)}
       + \dot \mu_{(i)} \alpha
       + \left( \mu_{(i)} + p_{(i)} \right) \kappa + \delta Q_{(i)}.
   \label{eq8}
\eea
{}From $T^{\;\;\;\; b}_{(i)\alpha;b} = Q_{(i)\alpha}$ we have
the momentum conservation of the fluid components
\bea
   \dot v_{(i)} + \left[ H \left( 1 - 3 c_{(i)}^2 \right)
       + { 1 + c_{(i)}^2 \over \mu_{(i)} + p_{(i)} } Q_{(i)} \right] v_{(i)}
       = {k \over a} \left[ \alpha
       + {1 \over \mu_{(i)} + p_{(i)}} \left( \delta p_{(i)}
       - {2 \over 3} {k^2 - 3 K \over a^2} \sigma_{(i)}
       - J_{(i)} \right) \right].
   \label{eq9}
\eea
By adding Eqs. (\ref{eq8},\ref{eq9}) over the components we have
Eqs. (\ref{eq6},\ref{eq7}).
In the multi-component situation 
Eqs. (\ref{eq1}-\ref{eq7},\ref{eq8},\ref{eq9}) are the complete set 
expressed in a gauge-ready form.
If we decompose the pressure as
$\delta p_{(i)} = c_{(i)}^2 \delta \mu_{(i)} + e_{(i)}$ with
$c_{(i)}^2 \equiv \dot p_{(i)}/\dot \mu_{(i)}$, comparing with
Eq. (\ref{p-decomposition}) we have
\bea
   e = \delta p - c_s^2 \delta \mu
       = \sum_{(l)} \left[ \left( c_{(l)}^2 - c_s^2 \right) \delta \mu_{(l)}
       + e_{(l)} \right].
   \label{e}
\eea

Under the gauge transformation, from Eq. (20) in \cite{PRW}, we have:
\bea
   & & \delta \tilde \mu_{(i)} = \delta \mu_{(i)} - \dot \mu_{(i)} \xi^t, \quad
       \delta \tilde p_{(i)} = \delta p_{(i)} - \dot p_{(i)} \xi^t, \quad
       \tilde v_{(i)} = v_{(i)} - {k \over a} \xi^t, 
   \nonumber \\
   & & \tilde J_{(i)} = J_{(i)} + Q_{(i)} \xi^t, \quad
       \delta \tilde Q_{(i)} = \delta Q_{(i)} - \dot Q_{(i)} \xi^t.
   \label{GT-2}
\eea
Thus, we have the following additional gauge conditions:
\bea
   \delta \mu_{(i)} \equiv 0, \quad
       \delta p_{(i)} \equiv 0, \quad
       v_{(i)}/k \equiv 0, \quad
       {\rm etc}.
   \label{Gauges-2}
\eea
Any one of these gauge conditions also fixes the temporal gauge condition
completely.
Studies of the multi-component situations
can be found in \cite{KS,PRW,KS-2}.

\subsection{Rotation and Gravitational Wave}
                                \label{sec:rot-GW}

{}For the vector-type perturbation, using Eqs. (\ref{metric},\ref{Tab}),
we have:
\bea
   & & 8 \pi G T^{(v)0}_{\;\;\;\;\;\alpha}
       = - {\Delta + 2 K \over 2 a^2} \left( B_\alpha + a \dot C_\alpha \right),
   \label{rot-1} \\
   & & 8 \pi G \delta T^{(v)\alpha}_{\;\;\;\;\;\beta}
       = {1 \over 2 a^3} \left\{ a^2 \left[ B^\alpha_{\;\;|\beta}
       + B_\beta^{\;\;|\alpha} + a \left( C^\alpha_{\;\;|\beta}
       + C_\beta^{\;\;|\alpha} \right)^\cdot \right] \right\}^\cdot,
   \label{rot-2} \\
   & & {1 \over a^3} \left( a^4 T^{(v)0}_{\;\;\;\;\;\alpha} \right)^\cdot
       = - \delta T^{(v)\beta}_{\;\;\;\;\;\alpha|\beta},
   \label{rot-3}
\eea
where Eq. (\ref{rot-3}) follows from $T^b_{\alpha ; b} = 0$, and 
apparently Eqs. (\ref{rot-1},\ref{rot-2}) follow from the Einstein equations.
We can show that Eq. (\ref{rot-3}) follows from Eqs. (\ref{rot-1},\ref{rot-2}).
We introduce a gauge-invariant velocity related variable 
$V^{(\omega)}_\alpha ({\bf x}, t)$ as 
$T^{(v)0}_{\;\;\;\;\;\alpha} \equiv (\mu + p) V^{(\omega)}_\alpha$.
We can show $V^{(\omega)}_\alpha \propto a \omega$ where $\omega$ 
is the amplitude of vorticity vector, see Sec. 5 of \cite{PRW}.
Thus, ignoring the anisotropic stress in the RHS of Eq. (\ref{rot-3})
we have the conservation of angular momentum as
\bea
   {\rm Angular \; Momentum} \sim a^3 \left( \mu + p \right)
       \times a \times V^{(\omega)}_\alpha ({\bf x}, t)
       = {\rm constant \; in \; time},
\eea
which is valid for general $K$, $\Lambda$, and $p(\mu)$ in general scale.

{}For the tensor-type perturbation, using Eqs. (\ref{metric},\ref{Tab})
in the Einstein equations, we have
\bea
   8 \pi G \delta T^{(t)\alpha}_{\;\;\;\;\;\beta}
       = \ddot C^\alpha_\beta + 3 H \dot C^\alpha_\beta
       - {\Delta - 2 K \over a^2} C^\alpha_\beta.
   \label{GW-eq}
\eea
In the large-scale limit, assuming $K = 0$ and ignoring the anisotropic
pressure Eq. (\ref{GW-eq}) has a general integral form solution
\bea
   C^\alpha_\beta ({\bf x}, t)
       = C^{\;\alpha}_{1\beta} ({\bf x}) - D^\alpha_\beta ({\bf x})
       \int^t {1 \over a^3} dt,
\eea
where $C^{\;\alpha}_{1\beta}$ and $D^\alpha_\beta$ are integration constants
for relatively growing and decaying solutions, respectively.
Thus, ignoring the transient term the amplitude of the gravitational
wave is conserved in the super-horizon scale considering general
evolution of the background equation of state.

The conservation of angular momentum and the equation for the
gravitational wave were first derived in \cite{Lifshitz}.
Evolutions in the multi-component situation were considered in
Sec. 5 of \cite{PRW}.

\section{Discussions}
                                \label{sec:Discussion}

We would like to make comments on related works in several {\it textbooks}:
Eq. (15.10.57) in \cite{Weinberg}, Eq. (10.118) in \cite{Peebles-2},
Eq. (11.5.2) in \cite{Coles}, Eq. (8.52) in \cite{Moss},
and problem 6.10 in \cite{Padmanabhan} are in error.
All these errors are essentially about the same point involved with
a fallible algebraic mistake in the synchronous gauge.
The correction in the case of \cite{Weinberg} was made in \cite{GRG}:
in a medium with a nonvanishing pressure the equation for the density 
fluctuation in the synchronous gauge becomes third order because of 
the presence of a gauge mode in addition to the physical growing 
and decaying solutions which is true even in the large-scale limit.
The truncated second order equation in \cite{Weinberg,Moss} 
picks up a gauge mode instead of the physical decaying solution in the 
synchronous gauge.
The errors in \cite{Peebles-2,Padmanabhan} are based on imposing 
the synchronous gauge and the comoving gauge simultaneously, and thus 
happen to end up with the same truncated second order equation 
as in \cite{Weinberg}.
In a medium with nonvanishing pressure one cannot impose the two gauge 
conditions simultaneously (even in the large-scale limit).  
In Sec. 11.5 of \cite{Coles} the authors proposed a simple modification
of the Newtonian theory which again happens to end up with the same incorrect 
equation as in the other books.
In the Appendix we elaborate our point.
For comments on other related errors in the literature, see Sec. 3.10 in 
\cite{Ideal} and \cite{GRG}.

We would like to remark that these errors found in the synchronous
gauge are not due to any esoteric aspect of the gauge choice in the 
relativistic perturbation analyses.
Although not fundamental, the number of errors found in the literature
concerning the synchronous gauge seem to indicate the importance of using 
the proper gauge condition in handling the problems. 
{}For our own choice of the preferred gauge-invariant variables 
suitable for handling the hydrodynamic perturbation and consequent analyses, 
see Sec. \ref{sec:Hydrodynamics}; the reasons for such choices are made in 
Sec. \ref{sec:Six-gauges}.
We wish to recall, however, that although one may need to do more algebra 
in tracing the remnant gauge mode, even the synchronous gauge condition
is adequate for handling many problems as was carefully done in the 
original study by Lifshitz in 1946 \cite{Lifshitz}.

In this paper we have tried to identify the variables in the relativistic
perturbation analysis which reproduce the correct Newtonian behavior
in the pressureless limit.
In the first part we have shown that
$\delta$ in the comoving gauge ($\delta_v$) and $v$ and $\varphi$ 
in the zero-shear gauge ($v_\chi$ and $\varphi_\chi$) 
show the same behavior as the corresponding Newtonian 
variables {\it in general scales}.
In fact, these results have already been presented in \cite{MDE}. 
Also, various general and asymptotic solutions for every variable
in the pressureless medium are presented in the Tables of \cite{MDE}.
Compared with \cite{MDE}, in Sec. \ref{sec:Six-gauges} we tried to reinforce 
the correspondence by explicitly showing the second order differential 
equations in several gauge conditions.
We have also added some additional insights gained after publishing \cite{MDE}.
The second part contains some original results.
Using the gauge-invariant variables $\delta_v$, $v_\chi$ and $\varphi_\chi$ 
we write the relativistic hydrodynamic cosmological perturbation equations in 
Eqs. (\ref{P-1}-\ref{P-4}); actually, these equations 
are also known in \cite{Bardeen-1980}.
The new results in this paper are the compact way of deriving the
general large-scale solutions in Eqs. (\ref{varphi_chi-sol}-\ref{v_chi-sol}), 
and the clarification of the general large-scale conservation property of
$\varphi_v$ in Eq. (\ref{varphi_v-sol}).

The underlying mathematical or physical reasons for 
the variables $\delta_v$, $\varphi_\chi$, $v_\chi$, and $\varphi_v$ 
having the distinguished behaviors, compared with many other available
gauge-invariant combinations, may still deserve further investigation.

\subsection*{Acknowledgments}

We thank Profs.  P. J. E. Peebles and A. M\'esz\'aros for useful discussions
and comments, and Prof. R. Brandenberger for interesting correspondences.
JH was supported by the KOSEF, Grant No. 95-0702-04-01-3 and
through the SRC program of SNU-CTP.
HN was supported by the DFG (Germany) and the KOSEF (Korea).

\section*{Appendix: Common errors in the synchronous gauge}

In the following we correct a minor confusion in the literature
concerning perturbation analyses in the synchronous gauge.
The argument is based on \cite{GRG}.
{}For simplicity, we consider a situation with $K = 0 = \Lambda$,
$e = 0 = \sigma$ and ${\rm w} = {\rm constant}$ (thus $c_s^2 = {\rm w}$).

The equation for the density perturbation in the {\it comoving gauge}
is given in Eq. (\ref{delta-eq}).
In our case we have
$$
   \ddot \delta_v + ( 2- 3 {\rm w}) H \dot \delta_v
      + \left[ c_s^2 {k^2 \over a^2}
      - 4 \pi G \mu (1 - {\rm w}) ( 1 + 3 {\rm w} ) \right] \delta_v = 0.
   \eqno{(A1)}
$$
The solution in the super-sound-horizon scale is presented in
Eq. (\ref{Sol-w=const}).

In the {\it synchronous gauge}, from Eqs. (\ref{eq5}-\ref{eq7}) we have
(the subindex $\alpha$ indicates the synchronous gauge variable)
$$
   \ddot \delta_\alpha + 2 H \dot \delta_\alpha
       + \left[ c_s^2 {k^2 \over a^2}
       - 4 \pi G \mu ( 1+ {\rm w} ) ( 1 + 3 {\rm w} ) \right] \delta_\alpha
       + 3 {\rm w} ( 1 + {\rm w}) {k \over a} H v_\alpha = 0,
   \eqno{(A2)}
$$
$$
   \dot v_\alpha + ( 1- 3 {\rm w}) H v_\alpha
       - {k \over a} {{\rm w} \over 1 + {\rm w}} \delta_\alpha = 0.
   \eqno{(A3)}
$$
Notice that for ${\rm w} \neq 0$ these two equations are generally coupled
even in the large-scale limit.
{}From these two equations we can derive a third order differential equation
for $\delta_\alpha$ as
$$
   \delta^{\cdot\cdot\cdot}_\alpha
       + {11 - 3 {\rm w} \over 2} H \ddot \delta_\alpha
       + \left( {5 - 24 {\rm w} - 9 {\rm w}^2 \over 2} H^2
       + c_s^2 {k^2 \over a^2} \right) \dot \delta_\alpha
$$
$$
       + \left[ {3 \over 4} \left( 1 + {\rm w} \right)
       \left( 1 + 3 {\rm w} \right) \left( -1 + 9 {\rm w} \right) H^3
       + {3 \over 2} \left( 1 + {\rm w} \right) H c_s^2 {k^2 \over a^2}
       \right] \delta_\alpha = 0.
   \eqno{(A4)}
$$
In the large-scale limit the solutions are
$$
   \delta_\alpha \quad \propto \quad
       t^{{ 2(1+ 3{\rm w}) \over 3(1 + {\rm w})}}, \quad
       t^{{9{\rm w} - 1 \over 3( 1+ {\rm w})}}, \quad
       t^{-1}.
   \eqno{(A5)}
$$
In Appendix B of \cite{GRG} we showed that
the third solution with $\delta_\alpha \propto t^{-1}$
is nothing but a gauge mode for a medium with ${\rm w} \neq 0$.
[As mentioned before, the combination $\delta_\alpha$ is not gauge-invariant.
 {}From Eq. (\ref{GT}) we have
 $\delta_\alpha \equiv \delta + 3 H ( 1 + {\rm w}) \int^t \alpha dt$
 and the lower bound of the integration gives rise to the gauge mode 
 which is proportional to $t^{-1}$.]
{}For a pressureless case the physical decaying solution also behaves 
as $t^{-1}$ and the second solution in Eq. (A5) is invalid,
see Sec. 4 in \cite{GRG}.
In Eq. (B5) of \cite{GRG} we derived the relation between solutions
in the two gauges explicitly; the growing solutions are the same in both gauges
whereas the decaying solutions differ by a factor
$(k / aH)^2 \propto t^{2(1 + 3 {\rm w}) \over 3 (1 + {\rm w})}$.

Now, we would like to point out that, in a medium with ${\rm w} \neq 0$,
one cannot ignore the last term in Eq. (A2)
even in the large-scale limit.
If we {\it incorrectly} ignore the last term in Eq. (A2) we recover
the {\it wrong equation} in the textbooks mentioned in 
Sec. \ref{sec:Discussion} which is
$$
   \ddot \delta_\alpha + 2 H \dot \delta_\alpha
       + \left[ c_s^2 {k^2 \over a^2}
       - 4 \pi G \mu ( 1 + {\rm w} ) ( 1 + 3 {\rm w} ) \right] \delta_\alpha
       = 0.
   \eqno{(A6)}
$$
In the large-scale we have solutions
$$
   \delta_\alpha \quad \propto \quad
       t^{{ 2(1+ 3{\rm w}) \over 3(1 + {\rm w})}}, \quad
       t^{-1}.
   \eqno{(A7)}
$$
By ignoring the last term in Eq. (A2), in the large-scale
limit we happen to recover the fictitious gauge mode under the price of losing
the physical decaying solution [the second solution in Eq. (A5)].
Thus, in the large-scale limit one cannot ignore the last term in Eq. (A2)
in a medium with a general pressure; the reason is obvious if we see
Eq. (A4).
Also, one cannot impose both the synchronous gauge condition and the
comoving gauge condition simultaneously.
If we simultaneously impose such two conditions,
thus setting $v \equiv 0 \equiv \alpha$, from Eq. (\ref{eq7}) we have
$$
   0 = {k \over a \left( 1 + {\rm w} \right) } \left( c_s^2 \delta
       + {e \over \mu} - {2\over 3} {k^2 - 3K \over a^2} {\sigma \over \mu}
       \right).
   \eqno{(A8)}
$$
Thus, for $e = 0 = \sigma$ and medium with the nonvanishing pressure
we have $\delta = 0$ which is a meaningless system;
this argument remains valid even in the large-scale limit.


\newpage

\noindent
{\bf Table 1.  Newtonian correspondences:}
{}For the synchronous gauge we ignore the gauge mode. 
Thus the synchronous gauge is equivalent to the comoving gauge.
X indicates that the behavior differs from the Newtonian one.
The small-scale implies the scale smaller than the (visual) horizon.
Explicit forms of exact and asymptotic solutions for every variable are 
presented in \cite{MDE,Ideal}.

\baselineskip=15pt

\noindent
=================================================
\begin{tabbing}
Gauge \hskip 4.7cm\= Variable \hskip 1.3cm \= General Scale \hskip 1cm
                                           \= Small Scale\\
---------------------------------------------------------------------------
--------------------------------------  \\
Comoving gauge          \> $\delta_v/k$        \> Newtonian \> Newtonian  \\
Zero-shear gauge        \> $\delta_\chi$       \> X         \> Newtonian  \\
Uniform-expansion gauge \> $\delta_\kappa$     \> X         \> Newtonian  \\
Synchronous gauge       \> $\delta_\alpha$     \> Newtonian \> Newtonian  \\
Uniform-curvature gauge \> $\delta_\varphi$    \> X         \> Newtonian  \\ 
Uniform-density gauge   \> $\delta \equiv 0$   \> 0         \> 0          \\
---------------------------------------------------------------------------
--------------------------------------  \\
Comoving gauge          \> $v \equiv 0$        \> 0         \> 0          \\
Zero-shear gauge        \> $v_\chi$            \> Newtonian \> Newtonian  \\
Uniform-expansion gauge \> $v_\kappa$          \> X         \> Newtonian  \\
Synchronous gauge       \> $v_\alpha$          \> 0         \> 0          \\
Uniform-curvature gauge \> $v_\varphi$         \> X         \> X          \\ 
Uniform-density gauge   \> $v_\delta$          \> X         \> X          \\
---------------------------------------------------------------------------
--------------------------------------  \\
Comoving gauge          \> $\varphi_v$         \> X         \> X          \\
Zero-shear gauge        \> $\varphi_\chi$      \> Newtonian \> Newtonian  \\
Uniform-expansion gauge \> $\varphi_\kappa$    \> X         \> Newtonian  \\
Synchronous gauge       \> $\varphi_\alpha$    \> X         \> X          \\
Uniform-curvature gauge \> $\varphi \equiv 0$  \> 0         \> 0          \\ 
Uniform-density gauge   \> $\varphi_\delta$    \> X         \> X          \\
---------------------------------------------------------------------------
--------------------------------------  \\
\end{tabbing}

\vskip .5cm
\noindent
{\bf Note added:}

We introduce a variable
$$
   \Phi \equiv \varphi_v - {K / a^2 \over 4 \pi G ( \mu + p)} \varphi_\chi.
   \eqno{(B.1)}
$$
{}From the general large-scale solutions in eqs. (47),(53) we can show
$$
   \Phi ({\bf x}, t) = C ({\bf x}),
   \eqno{(B.2)}
$$
where the dominating decaying solution vanished.
Thus, remarkably, $\Phi$ is conserved, in the limit of vanishing
$c_s^2 k^2 / a^2$ term, considering general $K$, $\Lambda$, and
time-varying $p(\mu)$.
In a pressureless medium eq. ($B.2$) is an exact solution valid 
in general scale.

\end{document}